\documentclass{aa}
\usepackage{graphicx,natbib,txfonts}

\begin{document}

   \title{AGB nucleosynthesis in the Large Magellanic Cloud\thanks{based on
          observations collected at the European Southern Observatory, Chile
          (programme 074.D-0619(A))}}
   \subtitle{Detailed abundance analysis of the RV Tauri star MACHO\,47.2496.8}

   \author{M. Reyniers\inst{1}\fnmsep\thanks{Postdoctoral fellow of the Fund for
          Scientific Research, Flanders}
          \and
          C. Abia
          \inst{2}
          \and
          H. Van Winckel
          \inst{1}
          \and
          T. Lloyd Evans
          \inst{3}
          \and
%          M. Lugaro
%          \inst{4}
%          \and
%          A. Bona{\v c}i\'c Marinovi\'c
%          \inst{4}
%          \and
          L. Decin
          \inst{1}\fnmsep$^{\star\star}$
          \and
          K. Eriksson
          \inst{4}
          \and K. R. Pollard
          \inst{5}
          }

   \offprints{M. Reyniers, maarten@ster.kuleuven.be}

   \institute{Instituut voor Sterrenkunde, Departement Natuurkunde en
              Sterrenkunde, K.U.Leuven, Celestijnenlaan 200D, 3001 Leuven,
              Belgium
         \and
         Dpto. F\'\i sica Te\'orica y del Cosmos, Universidad de Granada,
         18\,071, Granada, Spain
         \and
         SUPA, School of Physics and Astronomy, University of St. Andrews,
         North Haugh, St. Andrews, Fife KY16 9SS, Scotland, UK
         \and
%         Sterrenkundig Instituut, Universiteit Utrecht, P.O. Box 80\,000,
%         3508 TA Utrecht, The Netherlands
%         \and
         Department of Astronomy and Space Physics, Uppsala University,
         Box 515, 75120 Uppsala, Sweden
         \and
         Department of Physics and Astronomy, University of Canterbury,
         Private Bag 4800, Christchurch, New Zealand
         }

   \date{Received June 23, 2006; accepted October 2, 2006}

   \authorrunning{M. Reyniers et al.}
   \titlerunning{Abundance analysis of MACHO\,47.2496.8}

\abstract
{Abundance analysis of post-AGB objects as probes of AGB nucleosynthesis.}
{%In order to study the yields of the AGB
%nucleosynthesis in a more metal deficient environment than in
%the Galaxy,
A detailed photospheric abundance study is performed on
the carbon-rich post-AGB candidate MACHO\,47.2496.8 in the LMC.}
{High-resolution, high signal-to-noise ESO VLT-UVES spectra of MACHO\,47.2496.8
are analysed by performing detailed spectrum synthesis modelling using
state-of-the-art carbon-rich MARCS atmosphere models.}
{The spectrum of MACHO\,47.2496.8 is not only dominated by bands of carbon
bearing molecules, but also by lines of atomic transitions of s-process
elements. The metallicity of [Fe/H]\,=\,$-$1.4 is surprisingly low for a field
LMC star. The C/O ratio, however difficult to quantify, is greater than 2, and
the s-process enrichment is large: the light s-process elements are enhanced by
1.2\,dex compared to iron ([ls/Fe]\,$=$\,$+$1.2), while for the heavy s-process
elements an even stronger enrichment is measured: [hs/Fe]\,$=$\,$+$2.1. The
lead abundance is comparable to the [hs/Fe]. With its low intrinsic metallicity
and its luminosity at the low end of the carbon star luminosity function, the
star represents likely the final stage of a low initial mass star.}
{The LMC RV\,Tauri star MACHO\,47.2496.8 is highly carbon and s-process
enriched, and is most probable a genuine post-C(N-type) AGB star. This is the
first detailed abundance analysis of an extragalactic post-AGB star to date.}


   \maketitle

\section{Introduction}\label{sect:introdctn}
It is well known that the MACHO experiment towards the Large Magellanic Cloud
(LMC) yielded invaluable contributions to variable star research. The discovery
of five distinct period-luminosity (PL) relations of lower mass RGB and AGB
giants
\citep[e.g.][]{wood99} is probably the best known example. Another legacy of
the experiment, was the discovery of 33 Pop II Cepheids and RV Tauri stars in
the LMC \citep{alcock98} and the definition of a single PL relation of both
groups with the appearance of the RV\,Tauri characteristic lightcurves
(alternating deep and shallow minima) at the high luminosity end: the RV\,Tauri
stars are a direct extension of Pop II Cepheids to longer periods. Absolute
luminosities were derived and the RV\,Tauri stars were confirmed to be likely
post-AGB stars, making these objects members of the small group  of post-AGB
stars with known luminosities. For Galactic RV\,Tauri stars there is no direct
probe to the luminosity but they were identified as post-AGB stars by
\citet{jura86}, mainly on the basis of the presence of an IR excess due to
circumstellar dust around many of them.

The post-AGB phase is a short phase, so not many of these sources are known
\citep[see][for a catalogue of candidate post-AGB objects]{stasinska06}.
Their temperature-gravity domain makes it possible to study a very wide range
in chemical species by their atomic transitions (if optically bright enough).
They are therefore ideal objects to constrain the AGB nucleosynthetic and
evolutionary models. Interestingly, post-AGB stars are chemically much more
diverse than theoretically anticipated: only a very small group of objects
shows direct chemical evidence for AGB nucleosynthesis, being enhanced in
carbon and enhanced in neutron capture s-process elements
\citep[e.g. review by][and references therein]{vanwinckel03}. Since the
post-AGB evolutionary tracks pass the Cepheid instability strip, the pulsating
RV\,Tauri stars could be, in principle, good candidates to test further AGB
chemical evolutionary models.

In recent years it became, however, clear that also for RV Tauri stars the
chemical picture is complex and {\em chemical} evidence for a post-AGB nature
(possible C-enrichement and s-process overabundances) is {\em not} found in
Galactic RV\,Tauri stars \citep[except for maybe V453\,Oph, for which a mild
s-process overabundance was found, see][]{deroo05}. In RV\,Tauri stars,
depletion abundance patterns prevail \citep{giridhar05,maas05}. The basic
scenario of the badly understood depletion process involves a chemical
fractionation due to dust formation in the circumstellar environment followed
by a decoupling of the gas and the dust with a reaccretion of the cleaned gas
on the stellar photosphere, which leaves it depleted of the refractory
elements. \citet{waters92} showed that the most favourable circumstance for
this process to occur is, if the circumstellar dust is trapped in a disc. The
presence of a disc in evolved objects is likely to be related to binarity
\citep[e.g.][]{vanwinckel03} which means that binarity must be widespread in
RV\,Tauri stars \citep[see also][]{deruyter06}.

The LMC sample of RV\,Tauri stars with their known luminosity is a unique
sample to study the nature of these stars and of the post-AGB evolution in
general. Moreover, abundance analyses of post-3rd dredge-up stars in the LMC
should make it possible to study the yields of the AGB nucleosynthesis in a
more metal deficient environment than the Galaxy. With the advent of
high-resolution spectrographs on 8\,m class telescopes, it is now feasible to
study in detail the abundance patterns of these individual RV Tauri candidates
in the LMC. The brightest object of the
\citet{alcock98} sample, MACHO$^{*}$04:55:43.2-67:51:10, also named as
MACHO\,47.2496.8, was studied at low resolution by \citet{pollard00}
and by \citet{lloydevans04}.
The basic parameters of MACHO\,47.2496.8 are summarised in
Table~\ref{tab:smbdgegevens}. Contrary to what is found in Galactic RV\,Tauri
stars, this object shows clear indications of a strong C-enhancement and of
s-process overabundances. The low-resolution spectra at different photometric
phases of MACHO\,47.2496.8 show that it is  strongly carbon enriched
(C/O\,$>$\,1) with strong C$_2$ bands at deep minima. Spectra taken at a
resolution of 0.12\,nm with the 1.9m SAAO telescope around deep minimum (phase
0.88) and with the Anglo-Australian Telescope near maximum light (phase 0.61)
show that the C$_2$ bands disappear almost completely at maximum light, with
only the actual bandhead of the origin band at 5165\,\AA\ still readily
detectable. The spectrum at minimum showed strong bandheads of $^{12}$C$^{12}$C
at 4737 and 4715\,\AA\ but the 4744\,\AA\ bandhead of $^{12}$C$^{13}$C was very
weak or absent, indicative of a high $^{12}$C/$^{13}$C ratio. Enhancements of
the Ba\,{\sc ii} spectral features at 4554\,\AA\ and 4934\,\AA\ were detected
on these spectra, making this object a very interesting star to study the AGB
nucleosynthesis in the LMC in detail.

In this paper, we will present a detailed
abundance analysis based on high-resolution, high signal-to-noise VLT-UVES
spectra. The paper is organised as follows: in Sect.~\ref{sect:photomtr} we
summarise the available photometry of the object, and discuss its variability
and extinction. In Sect.~\ref{sect:uvesbsrvtns}, the high-resolution
observations are briefly discussed, while Sect.~\ref{sect:abundancenlss} is
devoted to the actual analysis and the abundance results. In
Sect.~\ref{sect:discussionntrxtr} we discuss the abundance pattern, and
address the question whether the star is intrinsically or extrinsically
enriched in s-process elements. We end with the main conclusions
(Sect.~\ref{sect:maincnclsns}).

\begin{table}\caption{Basic parameters of MACHO\,47.2496.8.}\label{tab:smbdgegevens}
\begin{center}
\begin{tabular}{rrr}
\hline\hline
\multicolumn{3}{c}{\rule[-0mm]{0mm}{4mm}\bf\large MACHO\,47.2496.8}\\[.4mm]
\hline
Coordinates   & $\alpha_{2000}$ &     4 55 43.23 \\
              & $\delta_{2000}$ & $-$67 51 10.4 \\
\hline
Mean magnitude   &   V &14.97 \\
\hline
Spectral type &  & Supergiant\\
              &  & pulsating between\\
              &  & early K and late F\\
\hline
\end{tabular}
\end{center}
{\small Source coordinates: MACHO; magnitude: \citet{alcock98}; spectral type:
\citet{pollard00}.}
\end{table}

\section{Variability, extinction and luminosity}\label{sect:photomtr}
\subsection{MACHO photometry}
The database of the MACHO project \citep{alcock92, alcock95}, accessible via
{\tt http://wwwmacho.anu.edu.au/}, contains 260 simultaneous measurements of
MACHO\,47.2496.8 in the blue MACHO filter V$_{\rm M}$ and red MACHO filter
R$_{\rm M}$, and 465 additional measurements in V$_{\rm M}$ only. We performed
a frequency analysis on the 725 V$_{\rm M}$ magnitudes, using the
Jurkevich-Stellingwerf phase dispersion minimization ({\sc pdm})
method \citep{stellingwerf78}, and deduced a ``formal'' period (the period
between two successive deep minima) of P\,$=$\,112.97\,d ($\theta$-statistic
of $\theta$\,$=$\,0.23), which is slightly longer than the period in
\citet{alcock98}, being P\,$=$\,112.45\,d. Two outliers were removed from
the frequency analysis. A part of the V$_{\rm M}$ light curve for
MACHO\,47.2496.8 is also shown in Fig.~1 of \citet{pollard00}. Our period of
112.97 days and the epoch of minimum light, JD2448673.384, are used as the
basis for the phases of observations reported here.

A realistic error estimate on a period determination is always difficult to
obtain. Here, we use a rather straightforward method to give a rough error
estimate on the period: the width of the {\sc pdm} minimum is a measure for the
spread on the ``acceptable'' periods, i.e. the periods for which the
corresponding phase diagrams are still acceptable. This width is
$\sim$\,3\,10$^{-5}$\,c/d in frequency space, resulting in an uncertainty
of $\pm$0.2\,d on the period. As an extra check, we performed an additional
frequency analysis on the 260 V$_{\rm M}$\,$-$\,R$_{\rm M}$ colour data points.
This analysis yielded a slightly shorter period (P\,=\,112.86\,d), but still
within the quoted error of $\pm$0.2\,d.

\begin{figure}
\resizebox*{\hsize}{!}{\includegraphics{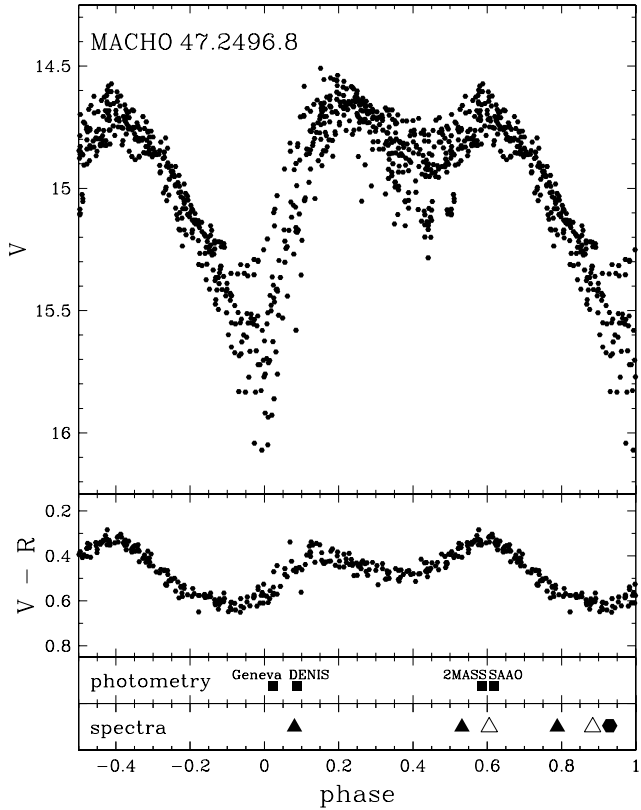}}
\caption{The MACHO V and V-R phase diagram, converted to the standard
Johnson V and Kron-Cousins R, constructed with a period of 112.97 days, and a
zero phase at deep minimum at JD2448673.384. For the V-R colour diagram, only
the simultaneous measurements in both bands are shown. At the bottom, the
phases of the photometric and spectroscopic data are given. The symbols of the
spectroscopic data are as follows: the full triangles are low resolution
spectra presented in \citet{pollard00}; the open ones are discussed in
\citet{lloydevans04}; the phase of the UVES spectrum (0.93) is shown with a
hexagon.}\label{fig:calibratedVphsdgrm}
% original location STER41/machocurve/phasediagrcalnewrev.ps
\end{figure}

The conversion of the raw MACHO magnitudes to a standard photometric system
is not straightforward, since the two filters are very broad, and they are only
partly overlapping with the classical Johnson or Kron-Cousins filters
\citep[see Fig.~1 of][for the filter pass bands]{alcock99}. The conversion to
the standard Johnson-Kron-Cousins system is performed using the formulae (1),
(2), (5) and (6) of \citet{alcock99}. In a first step, only the observations
that are taken simultaneously in both bands, were calibrated to the
Johnson-Kron-Cousins system, since V$_{\rm M}$\,$-$\,R$_{\rm M}$ is needed to
perform the conversion. Then, the calibration was extended to {\em all} 725
V$_{\rm M}$ points, using a calibration relation that was inferred from the
calibration of the 260 simultaneous points. A phase diagram of the converted V
magnitude, and the converted V-R colour, is shown in
Fig.~\ref{fig:calibratedVphsdgrm}.

From Fig.~\ref{fig:calibratedVphsdgrm} it is clear that the phase diagram
displays a large dispersion around the mean curve, especially around the
two minima. This dispersion is caused by cycle-to-cycle variations. Note that
the colour variations are much more stable over the covered cycles. Such
cycle-to-cycle behaviour in light and colour curves is often observed in
RV Tauri variables \citep[e.g.][]{pollard96}.

\subsection{Near-infared photometry}
Near-infrared photometry was obtained with the Mk III Infrared Photometer and
chopping secondary at the Cassegrain focus of the 1.9\,m reflector of the
South African Astronomical Observatory by one of us (KRP). This photometry
was complemented by near-IR survey photometry of 2MASS and DENIS. All near-IR
data, together with their JD and phase, are gathered in
Table~\ref{tab:allnearirdt}.
%By chance, all these measurements refer to the
%star when near maximum light.

\begin{table}
\caption{Optical Geneva and near-IR photometry of MACHO\,47.2496.8, together
with their JD and phase. The Geneva R and I are from the Gunn resp. Cousins
photometric systems.}\label{tab:allnearirdt}
\begin{center}
\begin{tabular}{rrrrrrrr}
\hline\hline
   JD     & Phase &   U   &  B    &   V    &   R  & I\\
\hline
\multicolumn{7}{c}{Geneva}\\
2453759.6 & 0.02 & 19.15 & 16.81 & 15.94 & 15.18 & 14.66 \\
\hline\hline
   JD     & Phase &   I   &  J    &   H    &   K  &\\
\hline
\multicolumn{7}{c}{SAAO}\\
2450437.4 &  0.62 &       & 13.18 &  12.67 &  12.54 &\\
2450438.3 &  0.62 &       & 13.20 &  12.56 &  12.29 &\\
Mean      &  0.62 &       & 13.19 &  12.61 &  12.41 &\\
$\sigma$  &       &       &   .04 &    .04 &    .05 &\\
\hline
\multicolumn{7}{c}{2MASS}\\
2451111.8 &  0.58 &       & 13.12 &  12.71 &  12.51 &\\
$\sigma$  &       &       &   .03 &    .04 &    .04 &\\
\hline
\multicolumn{7}{c}{DENIS}\\
2450377.8 &  0.09 & 14.02 & 13.22 &        &  12.53 &\\
$\sigma$  &       &   .03 &   .08 &        &    .13 &\\
\hline
\end{tabular}
\end{center}
{The Geneva R can be converted to R$_C$, the Cousins R magnitude, by the
conversion forumula from \citet{schombert90}, yielding R$_C$\,=\,14.89.}
\end{table}

\subsection{Optical CCD photometry with Euler}\label{sect:eulerpht}
Optical CCD photometry in five different bands was collected with the C2 camera
at the 1.2m Euler telescope on La Silla on 23/24 January, 2006. The filters that
were used are the Geneva U, B and V, completed with a Gunn-R and a Cousins-I
filter. Absolute photometry was derived by using exposures of the standard
stars HD\,47645 and SAO\,131547 that were taken immediately before and after
the MACHO\,47.2496.8 frames. This absolute calibration was cross-checked by the
extraction of two stars on the LMC-frame with published photometry which are
situated less than 4\,$\arcmin$ from MACHO\,47.2496.8: CSI-67-04552\,3
\citep[filters U, B and V,][]{schmidtkaler99} and MACHO\,47.2496.19
\citep[filters V and R,][]{keller02}. Taking into account that there are several
transformation formulae involved in the conversion of the Geneva and Gunn
filters to the Johnson-Cousins system \citep{schombert90, harmanec01}, the
magnitudes of the two check-stars are very close to the published ones.

Since there is a nearby star at approx. 2$\arcsec$, the extraction of
MACHO\,47.2496.8 was by no means easy. We corrected the pixels of
our target that are affected by this nearby star by their point symmetric
counterparts on the unaffected side of the photocenter. This induces an extra
uncertainty in the derived magnitudes, but, since the corrected pixels are
situated at the boundary of the psf, this uncertainty turned out to be not
larger than 0.05\,magn. The photometry is given in Table~\ref{tab:allnearirdt}.
The phase of the Euler observations, 0.02, has an estimated error of
$\sim$0.07, due to the quite large time gap between the MACHO photometry (on
which the period is based) and the C2-Euler observations. Note that the
MACHO photometry is not affected by this nearby star, since the latter star
has its own MACHO identification: MACHO\,47.2496.21.

\subsection{Spectral Energy Distribution}\label{subsect:sed}
\begin{figure}
\resizebox*{\hsize}{!}{\includegraphics{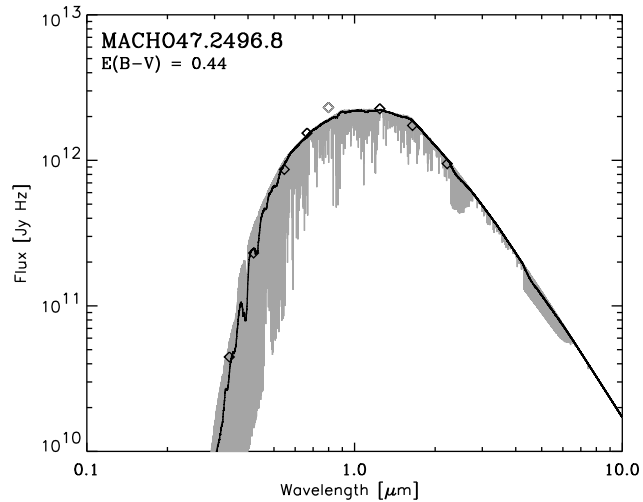}}
\caption{The spectral energy distribution (SED) of MACHO\,47.2496.8. Diamonds
are the measured magnitudes (converted to fluxes): Geneva U, B, V and Cousins
R taken with C2+Euler, I from DENIS (in gray), and SAAO J, H, K (see also
Table~\ref{tab:allnearirdt}). The MARCS model is shown in gray, while a smoothed
version is shown in a full black line. The minimisation is made using the
unsmoothed model in combination with the specific passbands of the photometric
filters involved. The I magnitude could not be fitted, and is excluded in the
minimisation. Possibly a phase difference is causing this discrepancy.}\label{fig:sedphtmtr}
% original location STER41/prbC2/voor_maarten/sed.ps
\end{figure}

The photometry from Table~\ref{tab:allnearirdt} was used to construct a Spectral
Energy Distribution of MACHO\,47.2496.8. The total reddening E(B-V) was
determined by minimising the difference between the magnitudes and a reddened
model atmosphere. For this model atmosphere, we used the parameters that were
determined in our abundance analysis (Sect.~\ref{subsect:atmosphprmtrs}). Note
that the optical Geneva photometry and the UVES spectrum from which the model
parameters are derived, differ less than 0.1 in phase. Such a small phase
difference is expected to result in only a small shift in atmospheric
parameters. The result can be found in Fig.~\ref{fig:sedphtmtr}; an
E(B-V)\,$=$\,0.44 was found. Due to the presence of the innumerable spectral
lines in the model atmosphere, the specific passbands of the filters used were
taken into account (Geneva system: {\tt http://obswww.unige.ch/}, DENIS:
\citealt{fouque00}, SAAO: \citealt{glass73}). The integration of the scaled
model combined with an LMC distance of 50\,kpc, yields a luminosity of
$\sim$5000\,L$_{\odot}$ (M$_{\rm bol}$\,$\simeq$\,$-$4.5).

A specific problem arose for the I magnitude. The Cousins I taken with Euler
turned out to be suspiciously low. The Cousins I was the only magnitude
that could not be checked with a star on the CCD frame itself (see
Sect.~\ref{sect:eulerpht}). The DENIS-I magnitude --in gray on
Fig.~\ref{fig:sedphtmtr}-- could not be reconciled with the other optical
photometry as well, possibly attributable to a phase shift. We did not take
either I magnitude into account in the minimisation. 

We made an additional SED with the MACHO photometry, confined to those
measurements with a a phase close to the UVES phase (difference $<$\,0.05),
since the model atmosphere parameters are based on the UVES spectrum. The
SED made with the MACHO V and R, completed with the DENIS I and the near-IR
JHK measurements, yielded a reddening of E(B-V)\,=\,0.33 and integration
of the scaled model yielded a total luminosity of $\sim$4600\,L$_{\odot}$
(M$_{\rm bol}$\,$\simeq$\,$-$4.4). The fit is, however, of a lower quality
than the fit in Fig.~\ref{fig:sedphtmtr}, hence we take the value of
5000\,L$_{\odot}$ as the final value for the luminosity of MACHO\,47.2496.8.

\subsection{Interstellar extinction}
The reddening derived in Sect.~\ref{subsect:sed}, is the {\em total} reddening,
and can have, besides the interstellar component, a circumstellar component as
well. The interstellar extinction towards the LMC is well studied in
literature. UBV photometry and spectral types of 450 OB stars in the western
half of the LMC
\citep{isserstedt75,isserstedt79,isserstedt82,ardeberg72, rousseau78,
crampton79, conti86,fitzpatrick88} give a mean E(B-V) of 0.13\,mag, using the
method and calibrations of \citet{fitzpatrick90}. There are larger values in OB
associations and a region of lower values in the North of the LMC. The region
surrounding MACHO\,47.2496.8 appears fairly uniform on the Digitised Sky Survey,
and the mean reddening of 16 stars within 30 arcmin is 0.11\,mag. However
\citet{zaritsky04} find that the distribution of reddening values for cool
stars differs from that of hot stars and is bimodal, such that stars have E(B-V)
near zero or near 0.15\,mag, depending on whether they lie in front of or behind
a widespread thin dust layer. We cannot place MACHO\,47.2496.8 or other carbon
stars relative to this dust layer and, therefore, have to assume that the
reddening is not significantly different from the average for carbon stars.

\section{High resolution observations}\label{sect:uvesbsrvtns}
High resolution, high signal-to-noise optical spectra of our programme star
were taken with the UVES spectrograph mounted on the VLT-UT2 (Kueyen)
telescope, in the program to obtain high-quality data of a
larger sample of post-AGB objects. The data were acquired in visitor
mode during ESO period \#74 by one of us (HVW). These observations fit in the
framework of our ongoing program to study the photospheric chemical composition
of stars in their last stages of evolution \citep[e.g.][]{reyniers04, deroo05}.
The resolving power of the UVES spectra varies between $\sim$60,000 and
$\sim$65,000. Spectra were taken with the chip unbinned. The spectral interval
covered and some other details about the observations are given in
Table~\ref{tab:observations}. Since there is another star at approx. 2\arcsec\
in the southwest of MACHO\,47.2496.8, the slit was placed perpendicular to this
direction. With this instrument setup, the contribution of the close
neighbour is negligible.

The reduction of our spectra was performed in the dedicated ``UVES context''
of the MIDAS environment and included bias correction, cosmic hit correction,
flat-fielding, background correction and sky correction. We used optimal
extraction to convert frames from pixel-pixel to pixel-order space. The spectra
were normalised by dividing the individual orders by a smoothed spline function
defined through interactively identified continuum points. For a detailed
description of the reduction procedure, we refer to \citet{reyniers02a}. In
Table~\ref{tab:observations}, we also list some indicative signal-to-noise
values of the final data product. Sample spectra can be found in
Figs.~\ref{fig:c2synth}, \ref{fig:zonderenmet} and \ref{fig:pbsnthss}. We note that the
most delicate step in the reduction procedure is the normalisation, especially
in the blue part of the spectrum, since the spectrum is so crowded that the
continuum is seldom reached. The continuum placement is the most important
source of uncertainty on the abundances derived from lines in this region.

\begin{table}\caption{Log of the high-resolution VLT-UVES observations. Small
spectral gaps occur between 5757\,\AA\ and 5833\,\AA\ and between 8521\,\AA\ and
8660\,\AA\ due to the spatial gap between the two UVES CCDs. Since continuum
spectral intervals are hardly found in the spectrum, the signal-to-noise ratios
S/N given in the rightmost column, are only indicative, and should be
interpreted more as lower limits than as fixed values.}\label{tab:observations}
\begin{center}
\begin{tabular}{ccccc}
\hline\hline
 date & UT    & exp.time & wavelength & S/N \\
      & start &  (sec)   & interval (\AA) &     \\
\hline
\multicolumn{5}{c}{VLT-UT2\,+\,UVES}\\
\hline
 2005-02-08 & 02:49 & 7200 & 4780$-$6808 & 70 \\ %RED580
 2005-02-09 & 00:30 & 7200 & 3758$-$4983 & 40 \\ %DI2 BLU437
 2005-02-09 & 00:30 & 7200 & 6705$-$10084 & 80 \\ %DI2 RED860
\hline
\end{tabular}
\end{center}
\end{table}
%BLU     3746.3 4981.5 |  374.5 - 498
%RED580L 4775   5772.5  | samen: 477.5 - 681
%RED580U 5831   6814.3  |
%RED860L 6705.5 8544.3 |  samen: 670.5 - 1055
%RED860U 8645.5 10550  |

\section{Abundance analysis}\label{sect:abundancenlss}
\subsection{Carbon rich model atmospheres}
Before the actual determination of the atmospheric parameters, a new grid of
models for cool carbon-rich stars was calculated with the MARCS code, using
opacity sampling in 11\,000 frequency points. Atomic, diatomic and polyatomic
(C$_2$H$_2$, HCN, and C$_3$) opacities were included in addition to the
continuous ones in the model atmosphere computations. The models are
hydrostatic, spherically symmetric and computed under the assumption of LTE;
convective energy transport is included using a Mixing Length Theory (MLT)
formulation. Further details on the carbon-star model atmospheres will be given
in \citet{jorgensen06}; the current grid of models is described by
\citet{gustafsson03}. The sub-grid used here was computed for a rather wide
range in T$_{\rm eff}$, log g, overall metallicity and C/O ratios; the stellar
mass was set to 1 M$_\odot$.

\subsection{Atmospheric parameters}\label{subsect:atmosphprmtrs}
The precise determination of the atmospheric parameters was by no means
straightforward. The main difficulty was the lack of existing comparison
spectra. Due to the high C/O ratio as inferred from the strong C$_2$ bands,
the spectrum of MACHO\,47.2496.8 mimics the spectrum of a cooler carbon star
in those spectral regions where molecular lines dominate. The atomic
transitions, however, indicate a significantly higher temperature than the
typical carbon AGB star temperatures (2500-3500\,K). The presence of the
molecular lines itself points to a temperature lower than the galactic F-type
s-process enriched post-AGB stars.

The final atmospheric parameters were derived by an iterative process of
fitting specific spectral regions and fine-tuning the parameters consecutively.
These specific regions are carefully chosen in the sense that they contain
spectral features strongly depending on the parameters, like lines of different
ionisation stages of the same element. Also the wings of the Balmer lines
H$\alpha$ and H$\beta$ were fitted. The final parameters that were adopted
further in the analysis are: T$_{\rm eff}$\,=\,4900\,K, $\log g$\,=\,0.0,
$\xi_t$\,=\,3.5\,km\,s$^{-1}$ and model metallicity [M/H]\,=\,$-$1.5. An
additional and independent check was made by an equivalent width study of
selected iron lines. The search for clean, unblended lines is not easy in
the crowded spectrum of MACHO\,47.2496.8, but a detailed study of 8
Fe\,{\sc i} lines and 5 Fe\,{\sc ii} nicely confirms the atmospheric
parameters as derived by the spectrum synthesis iteration.

The procedure of finding the atmospheric parameters of MACHO\,47.2496.8 was
also followed in previous analyses of very similar stars
\citep{abia02, delaverny06}. We refer to the detailed error analyses in those
papers for error estimates on the parameters. Here we only remind that
the typical errors are as follows: $\pm$250\,K on T$_{\rm eff}$, $\pm$\,0.5
on $\log g$, 1\,km\,s$^{-1}$ on the microturbulent velocity $\xi_t$ and
5\,km\,s$^{-1}$ on the macroturbulent velocity. The latter parameter does
not influence the line strengths, but only the line profiles, and is only
used to better match synthetic and observed spectra in the spectral syntheses.

\subsection{Spectral regions}\label{sect:spectralrgns}
We have analysed in full detail several spectral ranges from 4000\,\AA\ to
8000\,\AA. Since the molecular lines of the carbon bearing molecules are
comparable in strength to those in carbon stars, we particularly focussed our
spectral syntheses on those spectral regions that are also used in the analysis
of the cooler galactic carbon stars \citep[see e.g.][]{abia02,delaverny06}. For
these regions, line lists are carefully composed and these are discussed in
Section 3 of \citet{delaverny06} and will not be repeated here. In general,
however, the atomic data of many s-process transitions occuring in the whole
spectral region covered by our spectra, are not known to a good precision, which
prevents accurate quantitative abundance studies. The most important regions
are (i) between 4750\,\AA\ and 4950\,\AA\ for the s-process elements and the
mean metallicity, (ii) 6700-6730\,\AA\ for the Li line, (iii) 4050-4060\,\AA\ 
for Pb, (iv) 4260-4270\,\AA\ for Tc, (v) the C$_2$ band head around 4737\,\AA\ 
for the carbon abundance and isotopic ratio $^{12}$C/$^{13}$C (vi) the
CN complex between 8000 and 8100\,\AA\ for the N abundance and as an
additional tool for the carbon isotopic ratio. Theoretical spectra were
computed with the TurboSpectrum code
\citep{alvarez98, plez92a, plez92b, plez93} in spherical geometry. The
macroturbulence used ranges between 10 and 20\,km\,s$^{-1}$.

\begin{figure}
\resizebox*{\hsize}{!}{\includegraphics{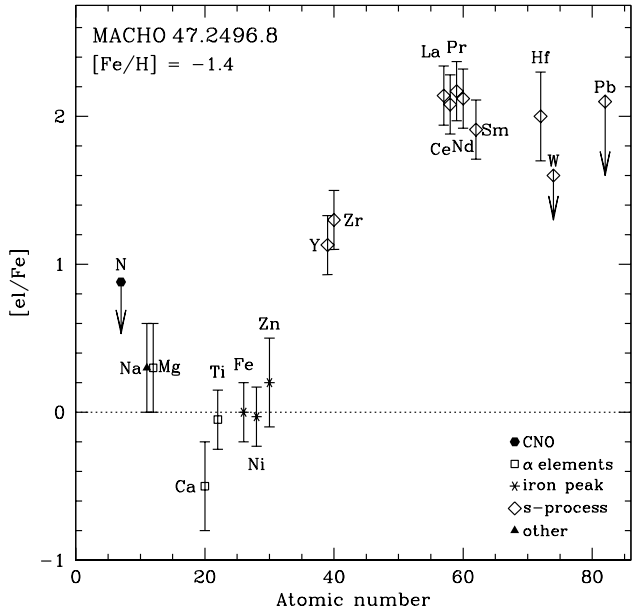}}
\caption{The abundances of MACHO\,47.2496.8 relative to iron [el/Fe].}\label{fig:elfevlsn_jan06}
% original location color STER41/alllmcstr/voorgranelfevlsn_jan06.ps
% original location STER41/analysislmcstr/alllmcstr/voorgranelfevlsn_jan06BW.ps
\end{figure}

The results of our abundance analysis are summarised in
Table~\ref{tab:absrslts}, and are also graphically presented in
Fig.~\ref{fig:elfevlsn_jan06}. The first column in Table~\ref{tab:absrslts}
gives the solar abundances $\log\epsilon_{\odot}$ of the elements that were
studied; the second one contains the actual ions; in the third one the method
of abundance determination is given (ss for spectrum synthesis and ew for the
equivalent width method); the fourth column gives the absolute abundances
derived $\log\epsilon$\,$=$\,$\log$(N(el)/N(H))+12; $\sigma_{\rm ltl}$, the
fifth column is the line-to-line scatter of the $n$ lines used (sixth column);
the last column gives the abundance relative to iron [el/Fe]. The solar
abundances needed to calculate the [el/Fe] values are taken from
\citet{grevesse98}, except: N \citep{hibbert91}, Mg \citep{holweger01}, La
\citep{lawler01}. Despite the fact that there are more recent values for some
of the solar abundances (especially for the solar CNO), we take these references
to ensure as much as possible consistency with our previous analyses, to be
able to compare abundance results adequately.

Errors on the reported abundances are difficult to assess, since very different
sources can contribute to the total error, and it is often very difficult
to perform an adequate error analysis. Main sources of error certainly include
continuum placement, undetected molecular and/or atomic features, an uncertain
effective temperature and inaccurate $\log(gf)$ values. For error estimates on
the reported abundances, we refer again to the error analyses in
\citet{abia02} and \citet{delaverny06}. In Fig.~\ref{fig:elfevlsn_jan06},
typical error bars are drawn, based on the number of lines used: if only one
line is used, an error of 0.3\,dex is applied, if more than one line is used,
an error of 0.2\,dex is assumed. The latter error estimate is justified if one
assumes that the error on an abundance based on $n$ lines, goes as
$\sqrt{\sum_i\sigma_i^2}/n$. In the following subsections, we will briefly
discuss our main findings for the different elemental groups.

\begin{table}
\caption{The abundance results for MACHO\,47.2496.8. The explanation of the
different columns is given in the text (Sect.~\ref{sect:spectralrgns}). At the
bottom, the usual s-process indices, together with metallicity, are given. More
information on the calculation of these indices is found in the
Sect.~\ref{sect:sprocesssbsctn}. The elements C and O are not separately 
included in the table due to the specific difficulties for these elements
that are discussed in Sect.~\ref{subsect:cno}.}\label{tab:absrslts}
\begin{center}
\begin{tabular}{rlcrrrr}
\hline\hline
\multicolumn{7}{c}{\rule[-0mm]{0mm}{5mm}{\Large\bf MACHO\,47.2496.8}}\\
\multicolumn{7}{c}{\bf MARCS model}\\
\multicolumn{7}{c}{\bf with main parameters}\\
\multicolumn{6}{c}{
$
\begin{array}{r@{\,=\,}l}
{\rm T}_{\rm eff} & 4900\,{\rm K}  \\
\log g  & 0.0\ {\rm(cgs)} \\
\xi_{\rm t} & 3.5\ {\rm km\,s}^{-1} \\
{\rm [M/H]} & -1.5 \\
\log(\epsilon({\rm C}) - \epsilon ({\rm O})) & 7.98
\end{array}
$
}&\\
\hline
$\log\epsilon_{\odot}$&ion&meth.&$\log\epsilon$&$\sigma_{\rm ltl}$&n&[el/Fe]\\
\hline
1.10 &Li\,{\sc i}   & ss &$<$1.5 &       & 1  &       \\
\hline
7.99 &N\,{\sc i}    & ss &$<$7.45&       &    &$<$0.88\\
\hline
6.33 &Na\,{\sc i}            & ss &  5.2  &       & 1  &  0.3  \\
7.54 &Mg\,{\sc i}            & ss &  6.4  &       & 1  &  0.3  \\
6.36 &Ca\,{\sc i}            & ss &  4.4  &       & 1  & $-$0.5  \\
5.02 &Ti\,{\sc ii}  & ss &  3.55 & 0.05  & 2  & $-$0.05 \\
7.51 &Fe\,{\sc i}   & ew &  5.98 & 0.10  & 8  & $-$0.11\\
7.51 &Fe\,{\sc ii}  & ew &  6.09 & 0.09  & 5  &       \\
6.25 &Ni\,{\sc i}            & ss &  4.8  & 0.00  & 2  &  0.0  \\
4.60 &Zn\,{\sc i}            & ss &  3.4  &       & 1  &  0.2  \\
\hline
2.24 &Y\,{\sc ii}   & ss &  1.95 & 0.1   & 4  &  1.13 \\
2.60 &Zr\,{\sc ii}  & ss &  2.5  &       & 1  &  1.3  \\
\hline
1.13 &La\,{\sc ii}  & ss &  1.85 & 0.09  & 9  &  2.14 \\
1.58 &Ce\,{\sc ii}  & ss &  2.24 & 0.10  &23  &  2.08 \\
0.71 &Pr\,{\sc ii}  & ss &  1.46 & 0.15  &12  &  2.17 \\
1.50 &Nd\,{\sc ii}  & ss &  2.20 & 0.09  &20  &  2.12 \\
1.01 &Sm\,{\sc ii}  & ss &  1.50 & 0.15  &12  &  1.91 \\
\hline
0.88 &Hf\,{\sc ii}  & ss &  1.5  &       & 1  &  2.0  \\
1.11 &W\,{\sc i}             & ss &$<$1.3 &       & 1  &$<$1.6 \\
1.95 &Pb\,{\sc i}            & ss &$<$2.6 &       & 1  &$<$2.1 \\
\hline
2.60 & Rb\,{\sc i}           & ss &\multicolumn{4}{l}{not detected} \\
\hline
 & & \multicolumn{2}{c}{\bf summary}& &\\
 &\multicolumn{2}{l}{[Fe/H]\,$=$\,$-$1.42}&\multicolumn{2}{l}{[hs/ls]\,$=$\,$+$0.90}&\\
 &\multicolumn{2}{l}{[ls/Fe]\,$=$\,$+$1.22}&\multicolumn{2}{l}{}&\\
 &\multicolumn{2}{l}{[hs/Fe]\,$=$\,$+$2.11}&\multicolumn{2}{l}{[Pb/hs]\,$\le$\,0.0}&\\
\hline
%\multicolumn{7}{l}{(d) dream loggfs}\\
%\multicolumn{7}{l}{(v) vald loggfs}\\
\end{tabular}
\end{center}
\end{table}

\subsection{CNO}\label{subsect:cno}
\begin{figure}
\resizebox*{\hsize}{!}{\includegraphics{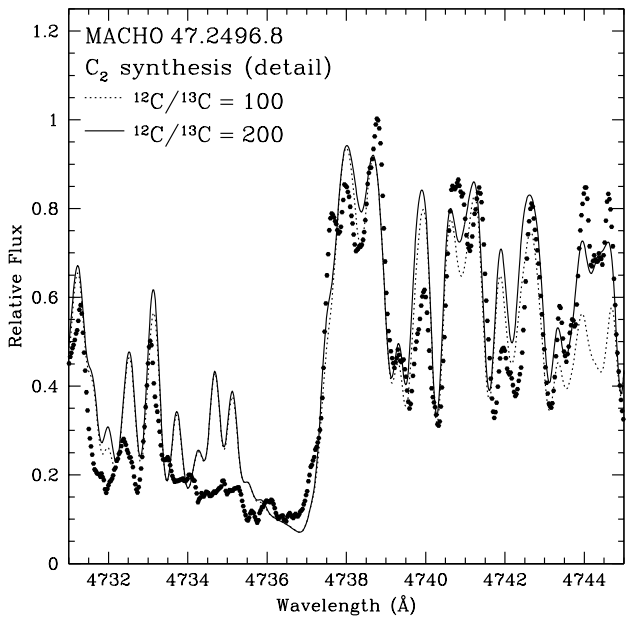}}
\caption{The VLT-UVES spectrum of MACHO\,47.2496.8 (points) around the C$_2$
4737\,\AA\ band head, overplotted with two spectrum syntheses with a different
$^{12}$C/$^{13}$C isotopic ratio.}\label{fig:c2synth}
% original location color STER41/granada/carlosspec/MACHO47_C2d.ps
% original location STER41/granada/carlosspec/MACHO47_C2dBW.ps
\end{figure}

Apart from the ``classical'' atmospheric parameters T$_{\rm eff}$ and $\log g$,
the C/O ratio or, more specifically, the difference between the C and O absolute abundances,
 is by far the most important ``extra'' parameter determining the
spectrum of a cool carbon-rich star. Unfortunately, it is impossible to
determine this difference accurately, since the O abundance cannot be determined
independently; spectral syntheses are very similar for different C and O
abundances, as long as the difference $\epsilon$(C)-$\epsilon$(O) is kept more
or less constant (even within 1 dex of variation of both abundances figures). The reason is that extra C and O atoms which  are added to the
photosphere in similar amount, immediately combine in the strong
CO-bond. Therefore, only a lower limit could be derived for the C/O ratio:
C/O\,$>$\,2. The carbon isotopic ratio is easier to determine and from the
synthesis of the band head region around 4737\,\AA\ (Fig.\ref{fig:c2synth}),
this ratio was found to be $^{12}$C/$^{13}$C\,$=$\,200\,$\pm$\,25. The error of
$\pm$\,25 indicates the formal error due to model parameters (T$_{\rm eff}$,
$\log g$, [M/H]), adding the problem of the continuum location and the
uncertainties in the absolute C and O abundances.

An upper limit for the nitrogen abundance is derived from the CN complex at
8000\,\AA, and is found to be N/H\,$<$\,2.8\,10$^{-5}$
($\log\epsilon(N)$\,$<$\,7.45). The spectrum synthesis of this region
provides also an additional test for the $^{12}$C/$^{13}$C ratio. The value
of 200 is compatible with the $^{12}$CN/$^{13}$CN features in the 8000\,\AA\ 
range. The derivation of the carbon isotopic ratio is, however, much more
difficult in this 8000\,\AA\ region, since the $^{13}$CN features are much
weaker than in the 4730\,\AA\ range, and the uncertainty of the fit is
therefore very large.

\subsection{Metallicity and $\alpha$-elements}
The metallicity was already roughly estimated in the iterative process of
the model atmosphere parameter determination. A careful synthesis of four clean
Fe\,{\sc i} lines yields $\log\epsilon$(Fe)\,=\,6.0, or [Fe/H]\,=\,$-$1.5.
We also calculated the abundance from 8 unblended Fe\,{\sc i} lines and 5
unblended Fe\,{\sc ii} lines through their equivalent width, and the results
agree nicely: $\log\epsilon$(Fe\,{\sc i})\,=\,5.98 and
$\log\epsilon$(Fe\,{\sc ii})\,=\,6.09. We will refer to the latter value as
``the'' metallicity of MACHO\,47.2496.8: [Fe/H]\,$=$\,$-$1.42. The iron
abundance derived from the Fe\,{\sc ii}-lines is chosen, since also at the
relatively low effective temperature of MACHO\,47.2496.8, the bulk of the iron
atoms in the photosphere is singly ionised. Note that the Fe proxies nickel
and zinc nicely follow the iron deficiency.

The $\alpha$-elements show surprisingly different abundances: Mg seems to be
enhanced by 0.3\,dex, while Ca is deficient by approximately 0.5\,dex. The
reason is not clear, but the difference is likely the initial composition of
the object without excluding an observational error due to blending: note that
we derive Mg and Ca abundances using only one line. The lighter
$\alpha$-elements, like Mg, are thought to be mainly produced during the
hydrostatic burning of massive stars, while the heavier $\alpha$-elements come
from explosive synthesis during SNe\,II. The abundances of the
$\alpha$-elements in MACHO\,47.2496.8 are in agreement with the observational
trends for the LMC that were recently derived by \citet{pompeia06}, so there is
no evidence for an intrinsic $\alpha$-enrichment or depletion. The same holds
for Na. Only Ca is somewhat lower than expected: \citet{pompeia06} predict
[Ca/Fe]$\sim$0 for a metallicity of [Fe/H]\,$=$\,$-$1.4, while we derived
[Ca/Fe]\,$=$\,$-$0.5 for MACHO\,47.2496.8.

\subsection{Lithium}
Although the region around the lithium doublet at 6708\,\AA\ is not very well
fitted, there is no evidence for an enhanced Li abundance. The upper limit
derived for Li is $\log\epsilon$(Li)\,$<$\,1.5. Note that the Li abundance
for a strongly s-process enriched object is difficult to determine, since the
Li doublet is strongly blended with a Ce\,{\sc ii} transition at 6708.10\,\AA\
\citep{reyniers02b}.

\subsection{S-process elements}\label{sect:sprocesssbsctn}
\begin{figure*}
\resizebox*{\hsize}{!}{\rotatebox{-90}{\includegraphics{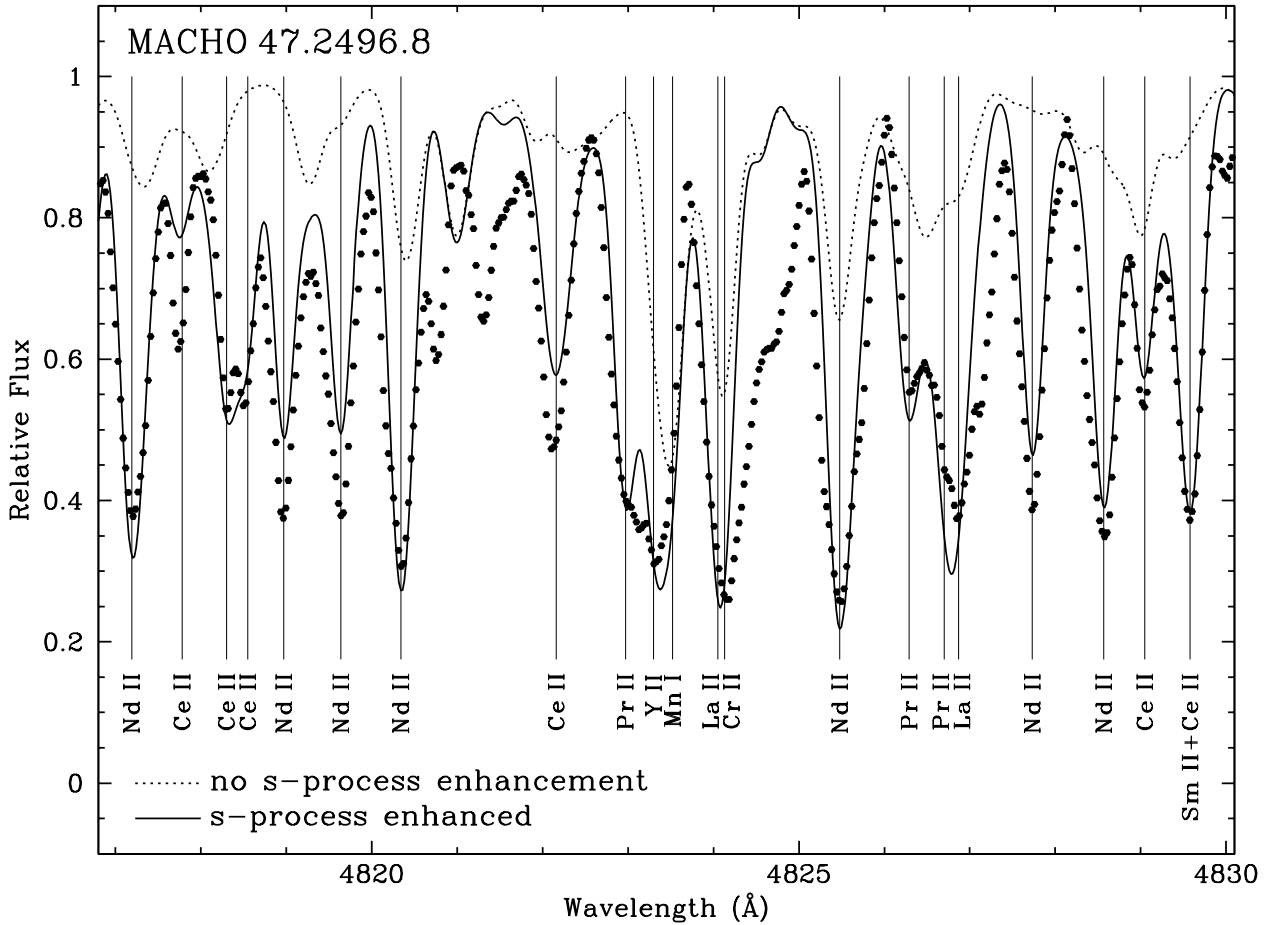}}}
\caption{The VLT-UVES spectrum of MACHO\,47.2496.8 (points) overplotted with two
spectrum syntheses. Both syntheses are made with the same model atmosphere,
only the abundances of the s-process elements differ. The synthesis in the
dotted line is made with solar abundances for the s-process elements (obviously
scaled down to the metallicity of MACHO\,47.2496.8), while the synthesis in the
full line is made with the enhanced s-process abundances as tabulated in
Table~\ref{tab:absrslts}. The observed spectrum is very well reproduced by the
spectrum synthesis, although there are clearly still some lines missing in the
line lists.}\label{fig:zonderenmet}
% original location color STER41/granada/carlosspec/MACHO47_sp2.ps
% original location STER41/granada/carlosspec/MACHO47_spd2BWid.ps
\end{figure*}

Although trace elements by nature, the absolute dominance of
transitions of s-process elements in the spectra is overwhelming.
This is illustrated in Fig.\ref{fig:zonderenmet}, in which
the synthesis is performed with and without s-process enhancement. S-process
elements are classically divided into the light ones noted ls (Sr, Y and Zr)
and the heavy ones, noted hs (Ba, La, Ce, Pr, Nd and Sm). Some elements show
a wealth of atomic transitions in the optical, like Ce and Nd, and the
abundances of these elements can be derived with great accuracy. Others, like
Sr and Ba, only produce a few strong resonance lines that are not suitable for
abundance determination purposes. This fact should be taken into account when
comparing individual s-process abundances. We could also derive an abundance
for the trans-Ba element hafnium (Hf), but it is based on only one line, so
it should be treated with caution.

We have also studied two resonance lines of Ba\,{\sc ii} to estimate the Ba
abundance. On top of being strongly saturated, the profiles are broader than
other atomic lines and  both lines show a blue-shifted component (14.6 and
15.5\,km\,s$^{-1}$, respectively). This is probably due to the dynamical
structure of the outer atmosphere of this pulsating star. Using only the
component at the same photospheric velocity as the other atomic lines, we
derive an upper limit of $\log\epsilon$(Ba)\,$<$\,2.8. We conclude, however,
that the special line profile makes the Ba abundance very uncertain. 

All s-process abundances are based on lines of singly ionised species.
Additionally, we have also searched for lines of doubly ionised Pr and Nd, which
are known to produce some strong lines in the spectra of enriched post-AGB stars
\citep[e.g.][]{reyniers04}. We found one Nd\,{\sc iii} line at 5294.099\,\AA\ 
and two lines of Pr\,{\sc iii} (at 5284.693\,\AA\ and 5998.930\,\AA) suitable
for abundance determination. The Nd\,{\sc iii} line yields the same
abundance as the one from the Nd\,{\sc ii} lines reported in
Table~\ref{tab:absrslts}; the two Pr\,{\sc iii} lines yield an abundance of
$\log\epsilon$(Pr)\,$=$\,1.85 and 2.26 respectively, and are much less
consistent with their singly ionised counterparts. It is not clear why
the discrepancy ($\sim$0.6\,dex) is so large.

In order to characterize the s-process pattern, and to be able to adequately
compare the observed pattern with other s-process enriched stars, three
s-process indices are usually defined: [ls/Fe], [hs/Fe] and [hs/ls].
Which specific elements are taken into account to determine these indices,
are, unfortunately, author-dependent, and are mainly just determined by the
elemental abundances that are obtained in the analysis. Here, we define the
ls-index as the mean of Y and Zr; the hs-index as the mean of La, Ce and Nd;
and obviously [hs/ls]$=$[hs/Fe]$-$[ls/Fe]. The elements Pr and Sm are excluded
from the hs-index to be able to compare with other s-process enriched stars.
Indeed, in many of the s-process studies, abundances for these elements are
lacking, and hence they are not incorporated in the indices. The resulting
indices are given at the bottom of Table~\ref{tab:absrslts}.

\subsection{Technetium}\label{sect:technetiumsbsctn}
The presence of technetium (Tc, Z\,=\,43) is a well known indicator that a star
is currently undergoing dredge-ups on the AGB phase, or that it has just left
this phase, since the half-life of $^{99}$Tc is $\sim$\,2$\times$\,10$^5$\,yr.
The detection of Tc is, however, complicated since all available lines are
quite weak, and they are often blended by much stronger atomic and/or molecular
lines. One of the lines that is often used in abundance studies of AGB stars,
is the resonance line at 4262.27\,\AA. The continuum placement is the main
source of uncertainty in this crowded spectral region. Moreover, at the
effective temperature of MACHO\,47.2496.8, the Tc line is intrinsically too weak
to see a clear difference in the synthesized line for different abundances.
We have to conclude that Tc cannot be used for MACHO\,47.2496.8 as a criterion
for the intrinsic enrichment.

\subsection{Niobium}\label{sect:niobiumsbsctn}
Besides Tc, there is another element whose abundance can be used as a criterion
to determine whether the star is intrinsically or extrinsically enriched:
niobium. Nb, which has only one stable isotope, is exclusively produced by the
decay of $^{93}$Zr (half-life $\tau_{1/2}\sim 1.5\times 10^6$ yr) and therefore
Nb populates the stellar surface only million years after the s-process elements
have been dredged-up. As a consequence, intrinsically (post-)AGB stars have
a Nb abundance much lower than extrinsically enriched stars for which
[Nb/Zr]\,$\simeq$\,0. For intrinsically enriched stars, the [Nb/Zr] abundance
depends only very slightly on the metallicity, and is predicted to be
[Nb/Zr]\,$\simeq$\,$-$1.0 \citep{bisterzo06}. For MACHO\,47.2496.8, we checked
all available Nb\,{\sc i} and Nb\,{\sc ii} lines in the spectrum, but
unfortunately none of them is suitable to derive a reliable abundance or upper
limit, since they are all situated in the blue part of the spectrum for which
the molecular veiling is most severe. There is one Nb\,{\sc ii} line at
4789.89\,\AA\ that could give an estimate for the upper limit. Again, the
continuum position is the main source of uncertainty in this spectral
region. Also, the blend in the immediate vicinity of the Nb
line at $\sim$4790.5\,\AA\ is clearly lacking in our line list. The derived
upper limit of [Nb/Zr]\,$\le$\,$-$0.3 is therefore very uncertain, but points
rather to an intrinsic than an extrinsic nature. Spectra taken in a hotter
phase might be more suitable to study this Nb line in more detail.

\subsection{Lead}\label{sect:leadsbsctn}
\begin{figure}
\resizebox*{\hsize}{!}{\includegraphics{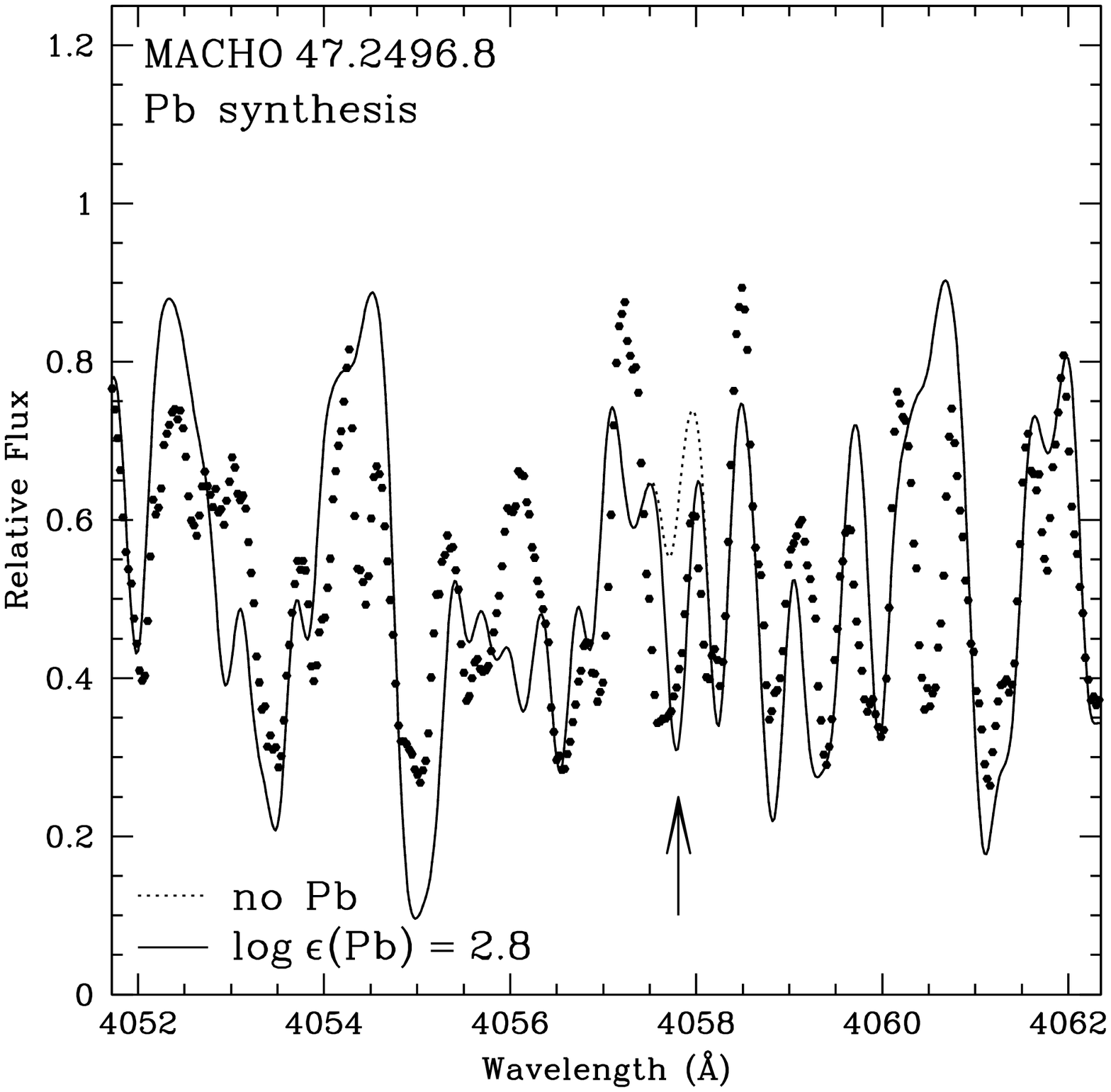}}
\caption{Synthesis of the Pb\,{\sc i} line at 4057.81\,\AA. The exact position
of the continuum is far from clear in the vicinity of the line, which makes it
only possible to derive an upper limit for the Pb abundance.}\label{fig:pbsnthss}
% original location STER41/granada/carlosspec/MACHO47_Pb.ps
% original location STER41/granada/carlosspec/MACHO47_PbBW.ps
\end{figure}

The end point of the s-process path is situated at lead and bismuth. Since the
discovery of the so-called ``lead stars'' by \citet{vaneck01}, lead has become
a very important issue in the study of the s-process. Unfortunately, lead
has only a few lines in the optical region, and they are all situated in
crowded regions in the blue spectral domain. In MACHO\,47.2496.8, only the line
at 4057.81\,\AA\ is accessible, and even there the blending by molecular lines
and s-process elements is severe. As a consequence, the main source of
uncertainty of the Pb synthesis in this region is the continuum placement.
Therefore, we can only derive an upper limit for the Pb abundance:
$\log\epsilon$(Pb)\,$<$\,2.8, as illustrated in Fig.~\ref{fig:pbsnthss}.

\section{Discussion: intrinsic or extrinsic?}\label{sect:discussionntrxtr}
Our abundance analysis (Table~\ref{tab:absrslts} and Fig.~\ref{fig:elfevlsn_jan06})
clearly shows that MACHO\,47.2496.8 is a metal poor, but strongly
s-process and carbon enriched object in the LMC. Contrary to Galactic
RV\,Tauri stars (see Sect.~\ref{sect:introdctn}),
there is no evidence for depletion of refractory
elements. In a carbon-rich (C/O$>1$) environment, ZrC is likely to condense
first \citep{lodders99} followed by TiC and SiC. For MACHO\,47.2496.8,
the abundances of these elements follow the trend which is expected from
a star that underwent an efficient dredge-up while evolving on the AGB,
so no depletion has taken place. To determine whether this enrichment is
intrinsic or extrinsic is not straightforward but in the following we argue
in favour for the intrinsic interpretation.

\subsection{Abundances}

From Sect.~\ref{sect:technetiumsbsctn} it is clear that the spectral feature,
Tc, that is classically used to determine whether a star is 
intrinsically or extrinsically enriched, cannot be used for MACHO\,47.2496.8,
mainly due to its relatively high T$_{\rm eff}$. An alternative indicator, the
[Nb/Zr] abundance discussed in \ref{sect:niobiumsbsctn}, is also difficult to
quantify, but might have some potential for spectra that are taken in a hotter
phase.

\subsection{Carbon star luminosity function}
Since carbon star counts in the Magellanic Clouds are considered to be nearly
complete, the luminosity functions for the carbon stars (CSLF) in the Clouds
can be constructed. In the LMC, 7750 carbon stars are detected, and a histogram
of the observational CSLF can be found in e.g. Fig.~6 of \citet{groenewegen06}.
This histogram does not indicate a clear faint limit to the luminosity function
of the N-type carbon stars.
The reproduction of the CSLF of the LMC by detailed nucleosynthetic models
is a long standing problem, since these models do not succeed in producing
the carbon stars at the low luminosity end of the CSLF. Recently, it has been
argued that the low-luminosity end of the CSLF is populated by first
giant-branch stars that are enriched by mass transfer from a former AGB
companion. This idea was quantitatively worked out in \citet{izzard04}. In
this paper, the transition between an extrinsically enriched carbon star and
a genuine AGB carbon star is situated around M$_{\rm bol}$\,$\sim$\,$-$4.
Consequently, the bolometric magnitude M$_{\rm bol}$\,$\simeq$\,$-$4.5 of
MACHO\,47.2496.8, inferred in Sect.~\ref{sect:photomtr}, does not provide
a strong constraint whether the object is either intrinsically or extrinsically
enriched. 
%Note however, that the bolometric correction might be not very accurate: 
%recently \citet{guandalini06} have shown that for genuine carbon stars, the
%bolometric luminosity is in many cases $\sim$1\,magn brighter when using IR
%photometry than the one inferred by the current calibrations.

We compared the flux of our program star to the flux of the N stars by
integrating over the wavebands B to K. The mean figure for MACHO\,47.2496.8
has been calculated assuming that the ratio between maximum and intensity-mean
light, calculated for V and R using the intensity-scaled light curves from the
MACHO website, applies to all wavelengths. This is a conservative assumption,
as the amplitude is probably less in the near infrared. The relative flux in
BVRI for the N stars has been assumed to be the same as for galactic N stars,
using the data of \citet{walker79}. Fluxes have been calculated in magnitudes
on an arbitrary scale, and show that MACHO\,47.2496.8 is 0.09 mag (at minimum
light) and 0.30 mag (at maximum light) brighter than the faint limit for N
stars in the LMC.

It is clear that MACHO\,47.2496.8 is located at the low luminosity end
of the carbon star luminosity function of the LMC. Its metallicity is,
however, quite low compared to the bulk of the LMC stars. A better
luminosity function may therefore come from the SMC where it is well
known that the CSLF is shifted to lower luminosities
\citep{groenewegen99}. The low intrinsic metallicity therefore
favours even more the intrinsic nature of the s-process enrichment.
In fact, at the metallicity of MACHO\,47.2496.8 ($\sim$ Z$_\odot/30$), 
the minimum luminosity for the formation of an AGB carbon star may be
slightly lower than M$_{\rm bol}\sim -4.0$, the exact value depending on
the mass-loss parametrization \citep[e.g.][]{straniero03}.

%{\tt\color[rgb]{0,1,0}If this star has lower metal content than that typical
%of LMC N-type carbon stars, then the appropriate lower limit in luminosity
%with which to make comparison may be considerably fainter than that of the
%LMC carbon stars. Compare the luminosity functions of LMC and SMC carbon stars
%according to Groenewegen (2004), also see the results of Demers et al,
%AJ, 123, 3428, 2002, and recall the greater ease of creating
%carbon stars on the AGB of a low metal content population. There are still
%theoretical difficulties, see Stancliffe et al in MN, 356, L1, 2005, but
%I prefer to follow the observations, and I doubt that all SMC carbon stars
%with M(bol) fainter than -4 can be CH stars, etc. In fact, even their successful
%prediction for the bottom of the N star distribution in the LMC is probably
%still too bright by 0.2 mag or so.}

\subsection{Comparison with galactic CH stars}\label{sect:CH}
CH-stars are metal poor, carbon rich giants with spectral types from G to K,
with enhanced CH, C$_2$ and s-process elements relative to normal giants.
CH-stars have typically also large radial velocities of
$\sim$100\,km\,s$^{-1}$. It is likely that all CH-stars are binaries in which
the present primary has gained processed material from a formerly-AGB companion,
now a white dwarf \citep{mcclure90}. The first accurate quantitative abundance
analyses were done by \citet{vanture92a, vanture92b, vanture92c}. He found that
the s-process abundance pattern of the CH-stars is characterised by large
s-process enhancements and large [hs/ls] ratios of $>$+0.9. Their metal
deficiency, carbon enrichment and s-process pattern are similar to
MACHO\,47.2496.8, and they are therefore ideal galactic comparison stars. The
majority of CH-giants, however, show very low carbon isotopic ratios, typically
$^{12}$C$/^{13}$C\,$<$\,10 \citep{vanture92a}, which is not seen in
MACHO\,47.2496.8.

More recent abundance analyses of CH-stars \citep{vaneck01, vaneck03, aoki01,
aoki02} confirmed the older results of \citet{vanture92c}, and focussed on the
detection of an enhanced lead content, which was predicted in different
nucleosynthetic AGB models \citep[e.g.][]{gallino98, goriely00}. As first
discovered by \citet{vaneck01}, a subgroup of the CH-stars does show this
predicted Pb enhancement (typically [Pb/hs]\,$\simeq$\,$+$1), but there are
also CH-stars that are not compatible with the predictions, showing [Pb/hs]
ratios that are more than one magnitude lower than predicted
\citep[see e.g. Fig.~4 in][]{vaneck03}.

The Pb abundance of MACHO\,47.2496.8 is very uncertain, and in
Sect.~\ref{sect:leadsbsctn}, we were only able to derive an upper limit of
[Pb/hs]$\sim$0. Therefore MACHO\,47.2496.8 is considered as a ``lead-low''
star. Nevertheless, the low [Pb/hs] ratio as well as the other s-process
indices are still compatible with recent nucleosynthesis calculations in 
low-mass AGB stars \citep{gallino06}. 

\section{Conclusion}\label{sect:maincnclsns}
In this paper, we have presented our study of the brightest RV Tauri star
that was found in the MACHO experiment: MACHO\,47.2496.8 \citep{alcock98}. The
star is strongly pulsating with a peak-to-peak amplitude in V of
$\sim$1.4\,magn.
%The reddening is therefore difficult to assess, since
%photometry in different systems is available, but not taken in the same
%pulsational phase.
A SED constructed using recent CCD photometry taken with the
Euler telescope at La Silla, indicates a reddening of E(B-V)\,=\,0.44, while the
interstellar reddening towards this region in the LMC is E(B-V)\,=\,0.11.
The intrinsic bolometric magnitude M$_{\rm bol}$ calculated from the spectral
energy distribution is $-$4.5, placing the star in the low luminosity end of
the carbon star luminosity function of the LMC.

Our high-resolution, high signal-to-noise optical UVES spectra are unfortunately
taken in a cool phase. The main results of our detailed abundance analysis
performed by a careful synthesis of some regions that are already well studied
in galactic carbon stars, are
\begin{itemize}
\item a surprisingly low metallicity: [Fe/H]\,$=$\,$-$1.4, with no evidence of
depletion
\item a high C/O ratio, certainly larger than 2, and a carbon isotopic ratio
$^{12}$C/$^{13}$C of around 200
\item a large s-process enrichment with [ls/Fe]\,$=$\,$+$1.2, and [hs/Fe]\,$=$\,$+$2.1
\item a low Pb abundance [Pb/Fe]\,$\simeq$\,[hs/Fe]
\end{itemize}
We used a new grid of state-of-the-art MARCS models for cool carbon rich stars,
which was especially constructed for this analysis.

Although we cannot totally exclude that the star is extrinsically enriched by a
former AGB companion which should be seen now as a white dwarf, the enrichment
is likely of an intrinsic origin. The luminosity, although quite low, is still
compatible with an advanced phase of a very low-mass star, and the pulsations
are prototypical of an RV Tauri like object. Unfortunately, the star is too hot
to use technetium as a criterion of an in situ enrichment. Also the alternative
element suitable as an intrinsic/extrinsic test, niobium, cannot be used for
this star. New spectra, taken in a hotter phase when the molecular veiling is
absent, could enable a conclusive test concerning the intrinsic/extrinsic
dilemma.

With its low luminosity and certainly very low inital metallicity,
MACHO\,47.2496.8 represents the final evolution of a star which must have had
an initial mass very close to solar. To our knowledge, this is the first
detailed chemical analysis of a post-AGB star with a known distance and
accurate luminosity estimate. The large carbon enhancement and the very rich
s-process nucleosynthesis show that also very low-mass objects will undergo
strong chemical changes during AGB evolution.

\begin{acknowledgements}
It is a pleasure to thank Maria Lugaro and Axel Bona{\v c}i\'c Marinovi\'c,
who were immediately willing to test their latest models on our results,
which initiated a most promising collaboration. The authors would also like to
thank Oscar Straniero for the Nb prediction, Pieter Deroo for the help with the
construction of the SED, Sophie Saesen for the Euler observations and the
anonymous referee for the many useful comments that improved the paper
considerably. The Geneva staff is thanked for observation time on the Euler
telescope. This paper utilizes public domain data obtained by the MACHO Project,
and data from the Vienna Atomic Line Database. MR and LD acknowledge financial
support from the Fund for Scientific Research - Flanders (Belgium); CA has been
partially supported by the Spanish grant AYA2005-08013-C03-03; KE thankfully
acknowledges support by the Swedish Research Council.
%ML gratefully acknowledges the support of NWO via the VENI grant.
\end{acknowledgements}

\bibliographystyle{aa}
%% the .bbl begins here

%% the .bbl ends here

\end{document}